\documentclass[aps,pra,onecolumn,superscriptaddress,nofootinbib,notitlepage]{revtex4-1}

\usepackage{graphicx}
\usepackage{mathptmx, textcomp}
\usepackage[usenames]{xcolor}
\usepackage{graphicx}  					
\usepackage{dcolumn}
\usepackage{bm}
\usepackage{amssymb, amsmath}
\usepackage[colorlinks,urlcolor=blue,citecolor=blue,linkcolor=blue]{hyperref}
\usepackage{cleveref}
\usepackage{units}

\graphicspath{{figures/}{../figures/final/}}

\newcommand{\B}{\textnormal{B}}
\newcommand{\aB}{a_{\B}}
\newcommand{\aI}{a}

\newcommand{\bra}[1]{\left<{#1}\right|}
\newcommand{\ket}[1]{\left|{#1}\right>}

\newcommand{\dif}{\textnormal{d}}
\newcommand{\voldif}[1]{\dif^{3}#1}

\renewcommand{\k}{{\bf k}}

\newcommand{\q}{{\bf q}}
\newcommand{\0}{{\bf 0}}

\newcommand{\nn}{\nonumber}

\def\ket#1{\left|#1\right\rangle}
\def\39K{$^{39}$K}
\def\87Rb{$^{87}$Rb}

\begin{document}

\title{Observation of Attractive and Repulsive Polarons in a Bose-Einstein Condensate}

\author{Nils B. J\o rgensen}
\affiliation{Institut for Fysik og Astronomi, Aarhus Universitet, 8000 Aarhus C, Denmark.}
\author{Lars Wacker}
\affiliation{Institut for Fysik og Astronomi, Aarhus Universitet, 8000 Aarhus C, Denmark.}
\author{Kristoffer T. Skalmstang}
\affiliation{Institut for Fysik og Astronomi, Aarhus Universitet, 8000 Aarhus C, Denmark.}
\author{Meera~M.~Parish}
\affiliation{School of Physics \& Astronomy, Monash University, Victoria 3800, Australia.}
\author{Jesper Levinsen}
\affiliation{School of Physics \& Astronomy, Monash University, Victoria 3800, Australia.}
\author{Rasmus S. Christensen}
\affiliation{Institut for Fysik og Astronomi, Aarhus Universitet, 8000 Aarhus C, Denmark.}
\author{Georg M. Bruun}
\affiliation{Institut for Fysik og Astronomi, Aarhus Universitet, 8000 Aarhus C, Denmark.}
\author{Jan J. Arlt}
\affiliation{Institut for Fysik og Astronomi, Aarhus Universitet, 8000 Aarhus C, Denmark.}

\date{\today}

\begin{abstract}
The problem of an impurity particle moving through a bosonic medium plays a fundamental role in physics. However, the canonical scenario of a mobile impurity immersed in a Bose-Einstein condensate (BEC) has not yet been realized. Here, we use radio frequency spectroscopy of ultracold bosonic $^{39}$K atoms to experimentally demonstrate the existence of a well-defined quasiparticle state of an impurity interacting with a BEC. We measure the energy of the impurity both for attractive and repulsive interactions, and find excellent agreement with theories that incorporate three-body correlations, both in the weak-coupling limits and across unitarity. The spectral response consists of a well-defined quasiparticle peak at weak coupling, while for increasing interaction strength, the spectrum is strongly broadened and becomes dominated by the many-body continuum of excited states. Crucially, no significant effects of three-body decay are observed. Our results open up exciting prospects for studying mobile impurities in a bosonic environment and strongly interacting Bose systems in general.
\end{abstract}

\maketitle

\twocolumngrid

The behavior of a mobile impurity interacting with its environment has provided deep insight into quantum many-body systems. Since Landau and Pekar first proposed that the coupling between electrons and lattice phonons leads to the existence of quasiparticles termed polarons~\cite{LandauPekar}, this idea has systematically been developed~\cite{Mahan2000book}.  The concept of the polaron is now central to our understanding of a wide range of systems, including technologically important semiconductors~\cite{Gershenson2006}, $^3$He--$^4$He mixtures~\cite{BaymPethick1991book}, and high temperature superconductors~\cite{Dagotto1994}. Indeed, even elementary particles of the Standard Model acquire their mass by coupling to the bosonic Higgs particle. 

Due to the great flexibility of atomic gases, the experimental realization of the impurity problem in a degenerate Fermi gas~\cite{Schirotzek2009,Kohstall2012,Koschorreck2012} has led to a dramatic improvement in our theoretical understanding of this fundamental problem \cite{Chevy2006,Prokofev2008,Mora2009,Punk2009,Combescot2009,Cui2010,Mathy2011,Massignan_Zaccanti_Bruun}. The bosonic counterpart has been subject of intense theoretical investigation~\cite{Cucchietti2006,Huang2009,Tempere2009,Rath2013,Li2014,Grusdt2014,Ardila2015,Christensen2015,Levinsen2015} and some specific cases have been studied experimentally: impurities interacting with an uncondensed bosonic medium~\cite{Spethmann2012}, charged or fixed impurities in a Bose-Einstein condensate (BEC)~\cite{Zipkes2010,Schmid2010,Balewski2013,Scelle2013,Marti2014}, and impurities confined to a lattice~\cite{Ospelkaus2006}. However, despite the importance of the impurity problem in a bosonic reservoir, there has not yet been a realization of the canonical mobile impurity in a BEC -- the Bose polaron. 

\begin{figure}[t]
	\centering
	\includegraphics[width=8.6cm]{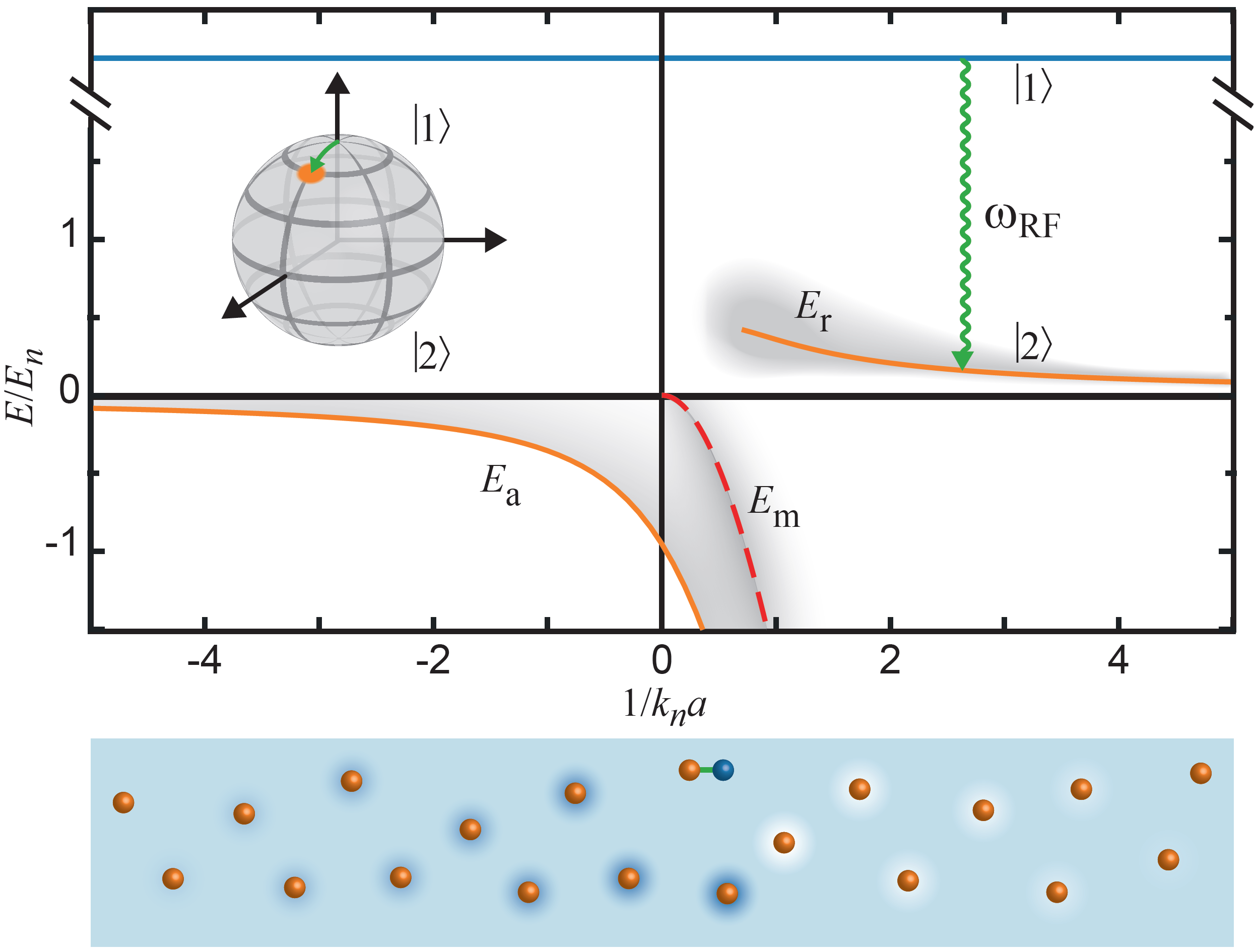}
	\caption{(color online). Sketch of the spectroscopic method and the impurity energy spectrum. A radio frequency pulse transfers atoms from the $\ket{1}$ to the $\ket{2}$ state. Only a small fraction is transferred, corresponding to a rotation by a small angle on the Bloch sphere (inset) in a non-interacting system. The solid lines show the energies of the zero-momentum attractive ($E_\text{a}$) and repulsive ($E_\text{r}$) polaron states in a uniform BEC as a function of the interaction parameter $1/k_n\aI$ (see text). The dashed line shows the molecular binding energy $E_\text{m}$ on the repulsive side of the Feshbach resonance, and the gray shading denotes a continuum of many-body states. The bottom cartoon shows impurity atoms (orange) in a BEC (blue); the intensity of the background color indicates the change in the BEC density due to the presence of impurity atoms.  
}
	\label{fig:setup}
\end{figure}

We investigate the Bose polaron using a harmonically trapped BEC of \39K atoms initially prepared in the $\ket{1} \equiv \ket{F=1,m_F=-1}$ state. To introduce impurities, we apply a radio frequency (RF) pulse, which transfers a small fraction of atoms into the $\ket{2} \equiv \ket{1,0}$ state, such that they can be regarded as isolated mobile impurities. The transition frequency $\omega_\text{RF}$ is changed from its unperturbed value $\omega_0$ due to the impurity-BEC interaction. This interaction, characterised by the $s$-wave scattering length $\aI$, is highly tuneable using a Feshbach resonance. Thus spectroscopy of the impurity state is performed, as shown schematically in Fig.~\ref{fig:setup}. We measure the energy of the impurity state and find a well-defined Bose polaron both for attractive and repulsive interactions, in good agreement with theories that incorporate three-body correlations. This constitutes the first observation of the Bose polaron for both attractive and repulsive interactions, in parallel with work at JILA~\cite{Hu2016}. 

Figure \ref{fig:setup} illustrates the behavior of the zero-momentum impurity in a uniform BEC. We parametrize the interaction strength as $k_na$ with the wavenumber $k_n=(6\pi^2n_0)^{1/3}$, where $n_0$ is the average BEC density. The energy scale of the system is $E_{n}=\hbar^2k_{n}^{2}/2m$, where $m$ is the mass of \39K. For weak interactions, $1/k_na\ll-1$ and $1/k_na\gg1$, the impurity forms well-defined quasiparticle states termed attractive and repulsive polaron, respectively. These have mean-field energy $4\pi\hbar^2 n_0\aI/m$, plus medium corrections which have recently been determined up to order $a^3$ \cite{Christensen2015}. On the attractive side of the Feshbach resonance, the zero-momentum attractive polaron is the ground state. In the absence of Efimov physics~\cite{Efimov1970,Braaten2006,Kraemer2006efe}, the attractive polaron state exhibits an avoided crossing with the molecular state beyond unitarity~\cite{Rath2013,Levinsen2015}. Above the ground state there is a continuum of many-body states, which in the weakly interacting limit is formed by polarons and Bogoliubov excitations with zero total momentum (Supplemental Material~\cite{SM}). On the repulsive side $1/k_na>0$, the polaron becomes increasingly damped when approaching the Feshbach resonance, since it can decay into a continuum of lower lying many-body states, and it is inherently metastable.

The experimental apparatus used to produce BECs is described in detail in~\cite{wacker2015}. Briefly, a dual-species magneto-optical trap captures \87Rb and \39K atoms and subsequently evaporative cooling is performed in a magnetic trap. All \87Rb atoms are evaporated leading to sympathetic cooling of \39K. The remaining \39K atoms are loaded into an optical dipole trap consisting of two crossed beams at a wavelength of $1064$nm. Two rapid adiabatic passages prepare the atoms in the $\ket{1}$ state by the transfers $\ket{2,2} \rightarrow \ket{2,-2}$ and $\ket{2,-2} \rightarrow \ket{1,-1}$. This path ensures that no atoms are produced in the $\ket{2}$ state due to a potentially imperfect state preparation. The sample is then further evaporatively cooled by lowering the dipole trap power. During this evaporation, a Feshbach resonance at $\unit[33.6]{G}$ is addressed to ensure efficient rethermalization. When a sufficiently low temperature is reached, the magnetic field is ramped to the desired value in the vicinity of the inter-state Feshbach resonance located at $\unit[113.8]{G}$~\cite{Lysebo2010,SM}. Finally, the power of the dipole trap is raised to increase the density of the BEC, which results in trap frequencies of $\nu_x = \unit[158]{Hz}$, $\nu_y = \unit[167]{Hz}$ and $\nu_z = \unit[228]{Hz}$. At this point, the BEC consists of $2\times 10^4$ atoms with average density $n_0=2.3\times10^{14}$ cm$^{-3}$ at a temperature $T = \unit[160]{nK}$ corresponding to $T/T_\text{c} \approx 0.6$ where $T_\text{c}$ is the critical temperature of Bose-Einstein condensation. The $\ket{1}$ atoms are weakly interacting with scattering length $\aB\approx9a_0$, where $a_0$ is the Bohr radius, such that $k_n\aB\approx0.01$. 

To form the polaron, a square RF pulse of $\unit[100]{\mu s}$ duration is used, which transfers a small fraction of atoms into the $\ket{2}$ state. This scheme of direct transfer ensures a perfect spatial overlap of impurities with the BEC. Furthermore, it is unique to a bosonic system since interaction effects in a Fermi gas subjected to a RF pulse are suppressed due to the Pauli principle~\cite{Zwierlein2003}. The pulse length and experimental magnetic field precision result in a spectral full width at half maximum of $0.15 E_n$. 

The RF transfer is described by the operator
$H_\text{RF}$ \hspace{0pt} $=$ $\Omega \text{e}^{-i\omega_\text{RF} t} \sum_{\mathbf k}a_{{\mathbf k}2}^\dagger a_{{\mathbf k}1}+\text{h.c.}$,
where $\Omega$ is the Rabi frequency. Within linear response, the resulting rate  of transfer into state $\ket{2}$ is given by 
$
\dot N_2=-2\Omega^2\text{Im}D(\omega_\text{RF})
$,
where $D(\omega_\text{RF})$ is the Fourier transform of the retarded spin-flip correlation function 
$D(t-t')=-i\theta(t-t')\langle[\sum_{\mathbf k}a_{{\mathbf k}1}^\dagger (t)a_{{\mathbf k}2}(t),\sum_{\mathbf k'}a_{{\mathbf k'}2}^\dagger (t') a_{{\mathbf k'}1}(t')]\rangle$. Since the BEC of  $|1\rangle$ atoms is weakly interacting, i.e., $n_0a_\text{B}^3\ll 1$, it can be described up to leading order in $n_0a_\text{B}^3$, yielding $D(\omega_\text{RF})=n_0G_2(\mathbf{k}=\mathbf{0},\omega_\text{RF})$ for a homogenous system,
where $G_2(\mathbf k,\omega_\text{RF})$ is the Green's function for an atom in spin-state $\ket{2}$ with momentum $\hbar \mathbf k$ and energy $\hbar \omega_\text{RF}$. It follows that an ideal RF measurement directly probes the $\mathbf k=\mathbf{0}$ part of the impurity spectral function, defined as $A(\omega_\text{RF})=-2\text{Im}G_2(\mathbf k=\mathbf{0},\omega_\text{RF})$. Note that we neglect vertex corrections to $D(\omega)$, which are small for $n_0a_\text{B}^3\ll 1$ as long as the temperature is much smaller than the critical temperature~\cite{Pethick2001}.

Subsequent to the RF pulse, the sample is held in the trap for a variable time before being released. After $\unit[5]{ms}$ of expansion, a strong magnetic field gradient is applied which separates the $\ket{1}$ and $\ket{2}$ components prior to absorption imaging after a total expansion time of $\unit[23]{ms}$. During hold time and expansion, three-body recombination processes involving two $\ket{1}$ and one $\ket{2}$ atom lead to a loss of atoms~\cite{Esry1999rot,Nielsen1999ler}. Due to the large atom number imbalance, all $\ket{2}$ atoms are typically lost at strong interactions and the number of $\ket{1}$ atoms is reduced accordingly. By performing a bimodal fit to the spatial distribution of $\ket{1}$ atoms the respective number of atoms in the BEC and thermal cloud are obtained. For strong interactions, we found that the number of lost atoms did not depend on the hold time in the trap and hence the three-body recombination processes during the initial expansion time is sufficient to fully remove $\ket{2}$ atoms. Thus, the hold time was set to zero for the majority of our measurements.

Due to the three-body recombination process, the number of lost atoms is three times larger than the number of atoms transferred to the polaron state. For each value of the interaction strength, we therefore choose the power of the RF pulse to provide a maximum loss of approximately $\unit[30]{} \%$ corresponding to a polaron fraction of $\unit[10]{} \%$. No significant deviations were observed by moderately varying the polaron fraction, which is discussed in detail in the Supplemental Material~\cite{SM}.

\begin{figure*}[htb]
	\centering
	\includegraphics[width=17.8cm]{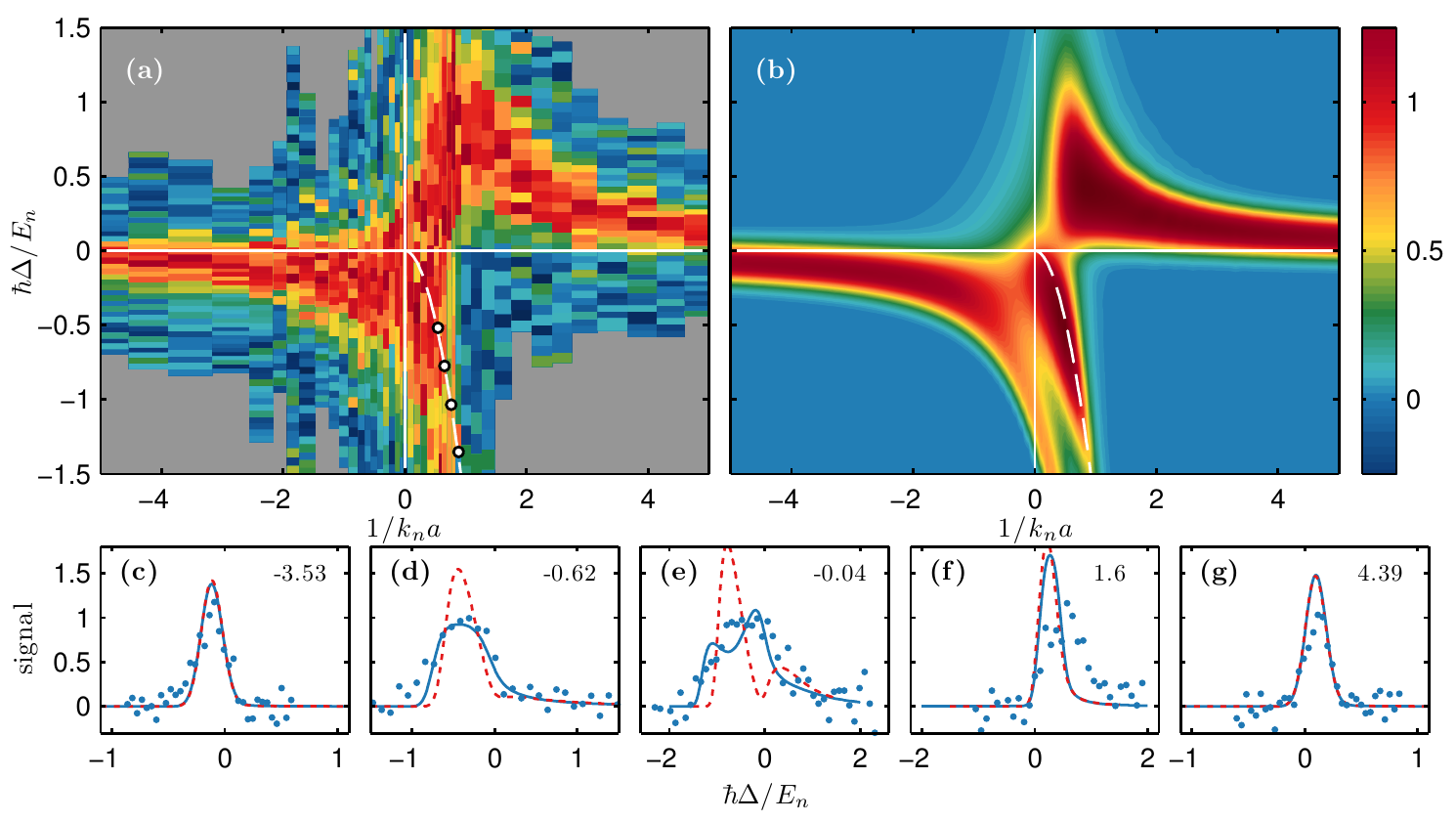}
	\caption{(color online). Spectral response of the impurity in the BEC. The false color plots show the experimentally measured spectroscopic signal (a) and the calculated spectrum (b), for different values of detuning $\Delta$ and the interaction parameter $1/k_n\aI$. The experimental spectrum is recorded such that its peak amplitude is constant for all values of $1/k_n\aI$. Accordingly, the theoretical spectrum is normalized such that its frequency integrated weight is the same as the experimental spectrum. In addition, the independently measured molecular binding energy (white dots) and a fit to it (dashed line) are shown~\cite{SM}. Negative values of the experimental signal are due to shot-to-shot atom number fluctuations. Panels (c)--(g) show the signal as a function of $\Delta$ for various values of $1/k_n\aI$ (see panel). The solid lines show the calculated signal, which  is in excellent agreement with the experiment, except for $1/k_n\aI=1.6$ where the agreement is qualitative. The dashed lines, obtained excluding three-body correlations, only agree with the experiment for weak interactions.}
	\label{fig:contour}
\end{figure*}

For a given interaction strength $k_n\aI$, we perform spectroscopy on the $\ket{1}\rightarrow\ket{2}$ transition by measuring the resulting BEC atom number in the $\ket{1}$ state as a function of detuning $\Delta=\omega_0-\omega_\text{RF}$, where $\omega_0$ is the unperturbed transition frequency between the two states~\cite{footnote1}.  Figure~\ref{fig:contour} compares the measured spectroscopic signal, corresponding to the normalized fraction of lost atoms, with that obtained by calculating the spectral function for a zero momentum impurity using a truncated basis method including three-body correlations (see Ref.~\cite{MJinprep} and the Supplemental Material~\cite{SM}). The theoretical calculation includes a spatial average over the trapped BEC and a convolution with the Fourier width of the RF pulse. It reproduces the observed signal strikingly well, both for attractive and repulsive interactions. In particular, both experiment and theory show a clear shift in the observed spectral weight due to the interaction between the impurity and the BEC. The calculation of the spectrum involves a restricted Hilbert space of impurity wavefunctions such that at most two Bogoliubov excitations of the BEC are included. Crucially, this truncated basis method (TBM) \cite{Cetina2016} allows us to include three-body correlations in the spectral function \textit{non-perturbatively}, and thus model the continuum of excited polaron states. Figure~\ref{fig:contour}(c)--(g) shows cuts through the spectrum at fixed $1/k_na$, demonstrating that the inclusion of three-body correlations  is essential for an accurate description of the strongly interacting unitary regime.

In contrast to the Fermi polaron~\cite{Schirotzek2009,Kohstall2012,Mora2009,Punk2009,Combescot2009,Prokofev2008,Massignan_Zaccanti_Bruun}, there is no sharp transition to a molecular state and the attractive polaron quasiparticle remains the ground state of the system for all interaction strengths. However, the spectral weight of the polaron is increasingly transferred to the continuum of higher-lying states as the strongly interacting unitary regime is approached from the attractive side of the Feshbach resonance. This feature is clearly apparent in both the observed and the calculated spectral response in Fig.~\ref{fig:contour}. For $1/k_n a > 0$, the structure of this continuum is determined by the molecular branch, and in the theoretical spectrum we see a clear suppression of spectral weight between the ground-state quasiparticle and the continuum. This is not apparent in the experimental spectrum, potentially due to atom number fluctuations or correlation effects not included in the theory. Significantly, the theory correctly captures the abrupt decrease in the observed signal at negative detuning for $1/k_na\gtrsim1$, where the molecule becomes deeply bound compared to $E_n$. The detailed comparison of spectroscopic signals in Fig.~\ref{fig:contour}~(c)-(e) further highlights the excellent agreement between theory and experiment for the attractive branch. 
 
\begin{figure}[t]
	\centering
	\includegraphics[width=8.6cm]{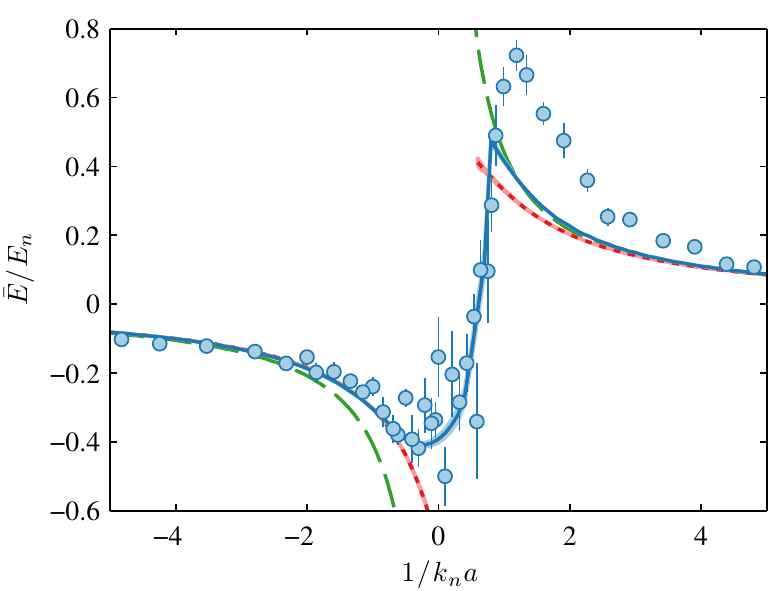}
	\caption{(color online). The average energy $\bar{E}$ of the impurity spectrum is shown as a function of the interaction parameter. The energy was obtained from Gaussian fits to the spectroscopic signal (blue dots) and to the full TBM spectrum (blue line). For comparison, we display the results for a TBM spectrum without three-body correlations (red line) and from perturbation theory (dashed line)~\cite{SM}.
	}
	\label{fig:fitPos}
\end{figure}

To further quantify the results, Fig.~\ref{fig:fitPos} compares the average impurity energy obtained from theory and experiment. For the attractive polaron, the experimental data agrees well with the results of the TBM. This holds even in the strongly interacting unitarity regime up to and including the abrupt shift of spectral weight to positive detuning at $1/k_na \simeq 1$. In the case of the repulsive polaron, the agreement is good for weak interaction, whereas there is only qualitative agreement for $1\lesssim1/k_n\aI\lesssim3$. This suggests that there are important aspects of the experiment that have not been included in the theory, such as effects of temperature, three-body recombination to deeply bound states, and multiple excitations of the BEC. The last effect is likely to play a role for strong interactions near $1/k_na \simeq 1$, since the repulsive branch in this regime involves a broad continuum of many-body states, which is challenging to model. For comparison, Fig.~\ref{fig:fitPos} also includes the result of the TBM without three-body correlations, highlighting the necessity of their inclusion.

Importantly, the perturbative result for the polaron energy~\cite{Christensen2015} accurately reproduces the observed energy shift in Fig.~\ref{fig:fitPos} for weak attractive and repulsive interactions. From this we conclude that the experimental data confirms the existence of a well-defined Bose polaron quasiparticle in this regime.

\begin{figure}[t]
	\centering
	\includegraphics[width=8.6cm]{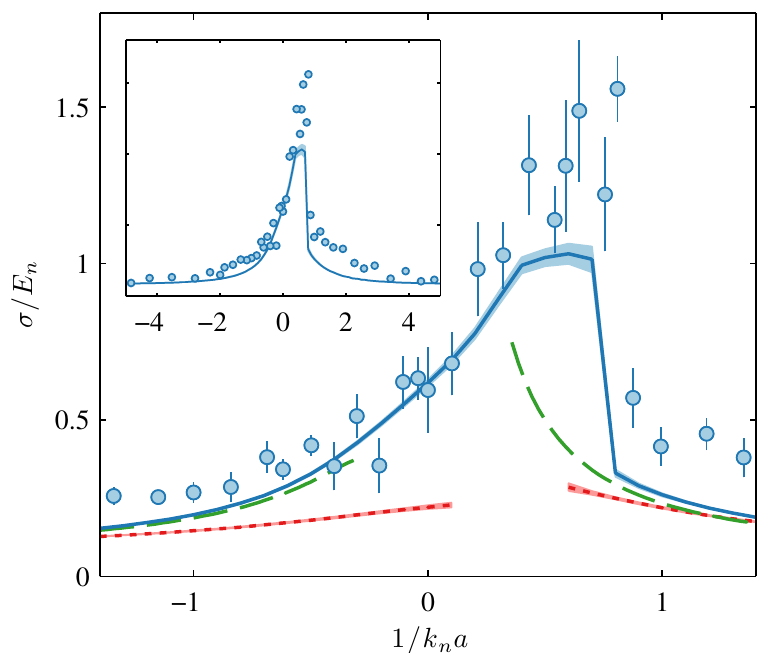}
	\caption{(color online). The width $\sigma$ of the impurity spectrum is shown as a function of the interaction parameter. The widths were obtained from Gaussian fits to the spectroscopic signal (blue dots), the full TBM spectrum (blue line), and the TBM without three-body correlations (red line). The green dashed line 
	was obtained from a spatial average and Fourier width convolution of the result from perturbation theory excluding the many-body continuum~\cite{SM}. The inset shows the width for the entire experimental data set compared with the full TBM spectrum.
}
	\label{fig:fitwidth}
\end{figure}

The width of the spectral response also agrees well with theory for interaction parameters $1/k_na \lesssim 1$ and $1/k_n a \gtrsim 3$, as shown in Fig.~\ref{fig:fitwidth}.  For weak interactions, the spectral broadening arises mainly from the  Fourier width of the RF pulse and the density inhomogeneity of the trapped BEC. This is illustrated by the fact that the perturbative result assuming a perfect undamped polaron reproduces the observed width in this regime~\cite{SM}.  However, near unitarity where the system is strongly correlated, the spectral weight of the polaron is small, and the many-body continuum of states accounts for the significant broadening of the spectrum. Importantly, this effect is captured by the TBM when three-body correlations are included. For the strongly interacting repulsive branch, there is again only qualitative agreement between theory and experiment for the reasons outlined above. 

Since strongly interacting Bose systems are expected to suffer from rapid three-body recombination, it is striking how well the experimental observations are described by theories that neglect such losses. The observed width of the spectrum is explained by the trap inhomogeneity, Fourier broadening, and the many-body continuum. 
Thus, while our method of detection relies on three-body losses, our results demonstrate that these processes occur on timescales longer than those associated with polaron physics. Therefore, the polaron is not significantly affected by three-body decay, indicating that it is long lived. We note that  the impurity decay rate, which is proportional to $n_0^2a^4$ when $n_0|a|^3\ll1$, is ultimately limited by the average interparticle spacing in the unitary regime. In this case the energy shift and decay rate both scale as $n_0^{2/3}$. Our results thus imply that the ratio of the decay rate to the energy shift at unitarity remains small, a finding which is consistent with the recent experiment on the unitary Bose gas~\cite{Makotyn2014}.

In conclusion, using RF spectroscopy on a BEC of $^{39}$K atoms, we  observed the existence of a long-lived Bose polaron. The spectral response is characterised by a well-defined quasiparticle peak both for weak attractive and repulsive interactions, whereas it becomes strongly broadened due to many-body effects for stronger interaction. We found excellent agreement with theoretical results including 3-body correlations, even for strong coupling. Our observation of a well-defined Bose polaron opens up the exciting opportunity to study quantum impurities in a bosonic environment systematically and in regimes never realized before. For instance, an intriguing question is how the polaron changes when the BEC melts. The effects of such a phase transition of the environment on an impurity particle has never been investigated before. Also, it will be interesting to examine the dynamics of the polaron and effects such as momentum relaxation, as well as induced interactions between  polarons and the formation of multipolaron states. There is also the prospect of observing stable Efimov trimers in a many-body environment for the first time~\cite{Nishida2015,Wei2015,Levinsen2015}. We do not expect to observe this in our present experiment, since the size of the smallest Efimov trimer is estimated to be 100 times larger than the interparticle spacing~\cite{Levinsen2015}. However, the Efimovian regime can be accessed by lowering the density or by using light impurities.

Note that parallel experiments were performed in a Bose-Fermi mixture, which report similar results~\cite{Hu2016}.

N. B. J., L. W., K. T. S. and J. J. A. acknowledge support from the Lundbeck Foundation and the Danish Council for Independent Research.
R. S. C. and G. M. B. acknowledge support from the Villum Foundation via Grant No. VKR023163.


\onecolumngrid

\section*{\large Supplementary Information: Observation of attractive and repulsive polarons in a Bose-Einstein condensate}

\section{Experimental details}
\subsection{Feshbach resonance structure}

The relevant interactions in our system are determined by two s-wave scattering lengths. The internal interaction of the BEC, which provides the bosonic medium, is governed by the scattering length $a_\text{B}$. The properties of the polaron are determined by the interaction between the impurity and the bosonic medium which is governed by the scattering length $a$.

\begin{figure}[b]
	\centering
	\includegraphics[width=8.9cm]{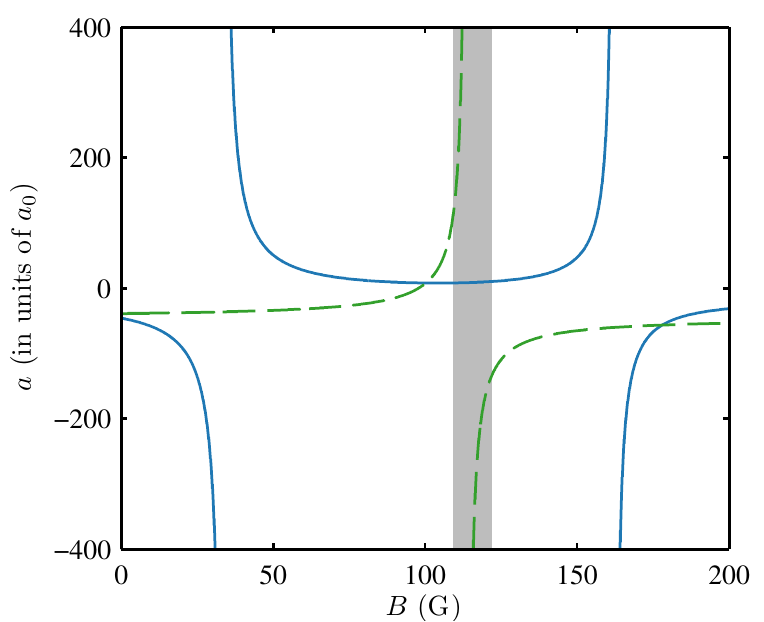}
	\caption{Feshbach resonance structure of the relevant states in \39K. The solid blue line shows the scattering length of atoms in the $\ket{1}$ state and the dashed green line shows the scattering length between atoms in the states $\ket{1}$ and $\ket{2}$. The shaded gray area displays the region in which measurements are performed.}
\label{fig:feshResStruct}
\end{figure}

The experiments are performed with a \39K BEC in the $\ket{1} \equiv \ket{F=1,m_F=-1}$ state and impurities in the $\ket{2} \equiv \ket{F=1,m_F=0}$ state. Three Feshbach resonances contribute to the two relevant scattering lengths as shown in Fig.~\ref{fig:feshResStruct}. The background scattering length of \39K is negative, but Feshbach resonances at $\unit[33.6]{G}$ and $\unit[162]{G}$ create a wide window of positive scattering length which allows stable BEC formation~\cite{derrico2007,wacker2015}. Within this window, an interstate Feshbach resonance allows us to tune the interactions between the impurity and the medium~\cite{Lysebo2010}. In the region where measurements are performed, the scattering length of the medium $a_\text{B}$ is approximately constant at $9a_0$. The scattering length which characterizes the interaction between atoms in the $\ket{2}$ state is approximately $-20 a_0$ in this region.

\begin{figure*}[htb]
	\centering
	\includegraphics[width=16.5cm]{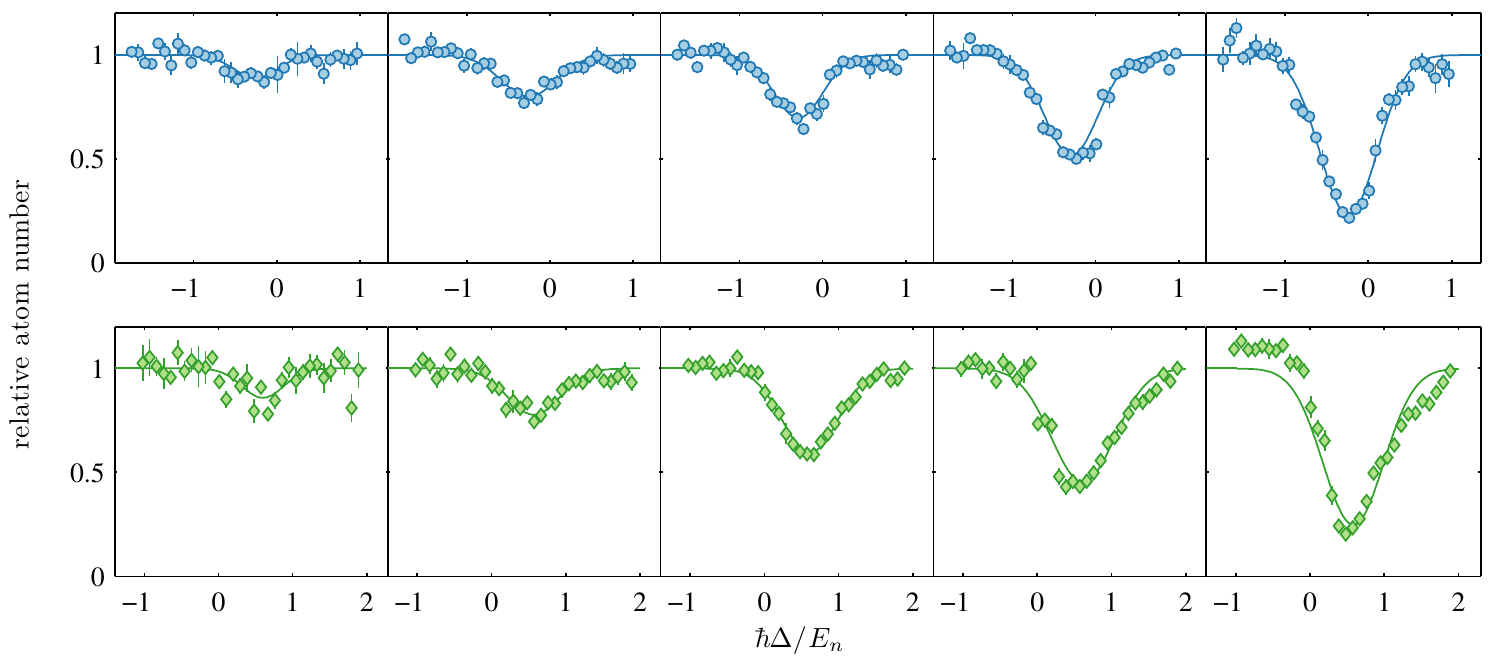}
	\includegraphics[width=16.5cm]{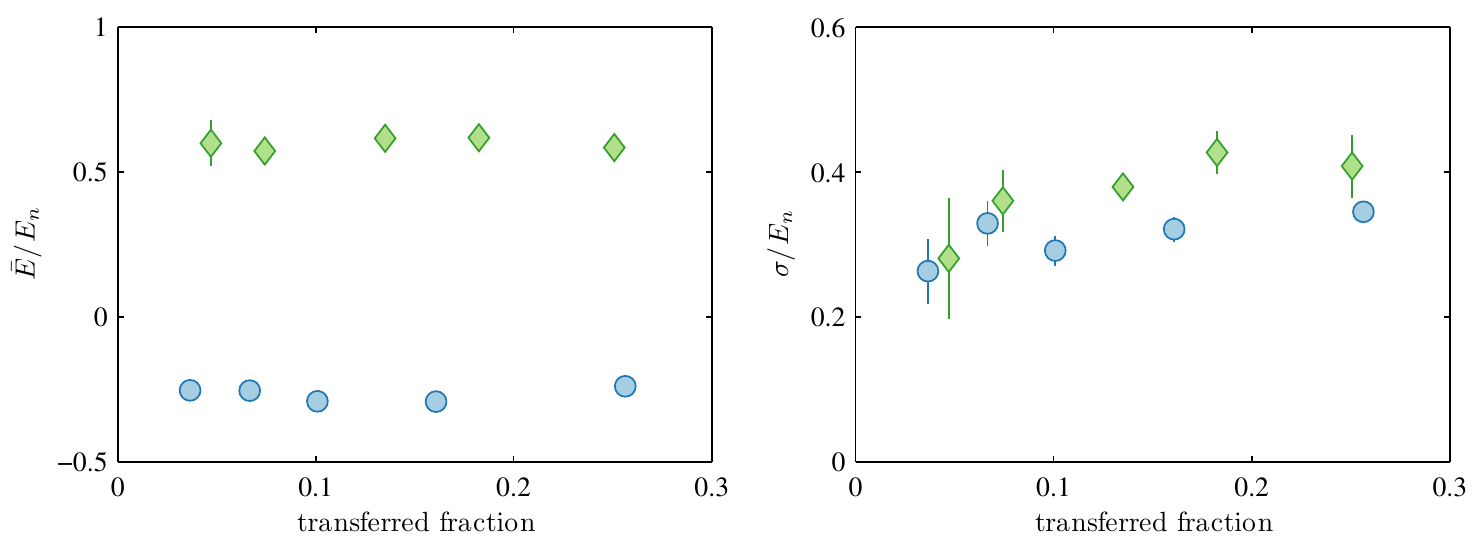}
	\caption{Polaron signal at various transferred fractions. Remaining number of atoms in the $\ket{1}$ state for $1/k_na = -0.84$ (top row) and $1/k_na = 1.6$ (middle row) as
	a function of the detuning for various RF powers. Bottom: Average energy $\bar{E}$ and width $\sigma$ of the spectroscopic signal as a function of transferred fraction for $1/k_na = -0.84$ (circles) and $1/k_na = 1.6$ (diamonds). The transferred fraction corresponds to a third of the relative loss due to three-body recombination.}
\label{fig:polaronFraction}
\end{figure*}

\subsection{Polaron fraction}

Polarons are formed by applying a RF pulse close to the transition between the $\ket{1}$ and $\ket{2}$ states. A fraction of $\ket{1}$ atoms is transferred to the $\ket{2}$ state and, due to the interaction, polarons are formed (see Sec.~\ref{TheorySec}). This method breaks down for large transferred fractions of atoms, since atoms in the $\ket{2}$ state cannot be treated as impurities in this case, and the bosonic medium of $\ket{1}$ atoms is depleted, thus changing its properties.

To investigate possible consequences of these effects, the spectroscopic signal was recorded for various transferred fractions at two values of the interaction parameter $1/k_na = -0.84$ and $1/k_na = 1.6$. To vary the transferred fraction, the power of the RF pulse was changed, while keeping the pulse duration constant. The observed signal as well as the resulting average energy and width are shown in Fig.~\ref{fig:polaronFraction}. 

No significant deviations of the average energy and only a minor increase in width are observed as a function of the transferred fraction. For transferred fractions beyond 15\%, a small distortion of the line shape is observed. We attribute this to a combination of BEC depletion, power broadening effects and a non-linear response of the broad many-body continuum part of the spectrum. 

Since these effects only appear at transferred fractions well above 10\%, it confirms that our measurements are performed within the linear response regime and represent a valid characterization of the polaron.


\section{Theoretical description}
\label{TheorySec}

In this section, we describe the theoretical framework used to interpret the experimental results. We consider the case of an impurity in a uniform BEC with density $n_0=2.3 \times \unit[10^{14}]{cm}^{-3}$ at zero temperature.  In the next section, we include the trap inhomogeneity using the local density approximation. We take $\aB=9a_0$  in all theory calculations, such that $k_n\aB\approx0.01$ and the BEC is weakly interacting. In the following, we set $\hbar$ and the volume to 1.

\subsection{Model of the Feshbach resonance in a $^{39}$K BEC}
We model the interactions between the $\ket{2}$  impurity atoms and a BEC of $\ket{1}$ atoms using a two-channel Hamiltonian for the Feshbach resonance. Within Bogoliubov mean-field theory for the condensed atoms, we have:
\begin{align}
  \hat{H} = & \sum_\k \left[ E_\k \beta^\dag_\k \beta_\k +
              \epsilon_\k
              a^\dag_{\k,2} a_{\k,2} + 
              \left( \epsilon^{\rm d}_\k + \nu_0 \right)
              d^\dag_\k d_\k \right ] \nn  \\ & \hspace{30mm}
+g \sqrt{n_0} \sum_\k \left( d^\dag_\k a_{\k,2}+ h.c. \right)
+ g \sum_{\k,\q } \left(d^\dag_\q 
  a_{\q -\k,2}
a_{\k,1} + h.c.   \right).
\label{eq:Ham2}
\end{align}
Here, $a_{{\mathbf k}\sigma}$ removes a $^{39}$K atom in spin state $\ket{\sigma}$ with momentum $\mathbf k$ and single-particle energy $\epsilon_\k=k^2/2m$; $m$ is the mass of the atom, $E_\k = \sqrt{\epsilon_\k (\epsilon_\k + 2\mu)}$ is the Bogoliubov dispersion, and $\mu = 4\pi \aB n_0/m$ is the chemical potential of the BEC, where  $\aB$ is the scattering length between the $|1\rangle$ atoms. 
The annihilation operator of $\ket{1}$ atoms is related to the creation and annihilation operators of Bogoliubov modes,
$\beta^\dagger_\k$ and $\beta_\k$ respectively, through
$a_{\k,1}=u_\k\beta_\k -v_\k\beta^\dag_{-\k}$. The coherence factors, given by
$u_\k^2 = [1 + (\epsilon_\k + \mu)/E_\k]/2$ and
$v_\k^2 = [-1 + (\epsilon_\k + \mu)/E_\k]/2$, are real and
positive. The atoms in the two spin states interact via a closed channel molecule. This has creation operator operator $d^\dagger_\k$ at momentum $\k$, single particle energy $\epsilon_\k^\text{d}=\epsilon_\k/2$, and a detuning $\nu_0$ from the two-atom $\ket{1}$-$\ket{2}$ threshold. The strength of the interaction is denoted $g$, and it is taken to be constant for momenta $|\k|<\Lambda$ and is set to 0 above the momentum cutoff $\Lambda$. Renormalization of the $\ket{1}$-$\ket{2}$ two-body  interaction then yields, respectively, the scattering length and range parameter~\cite{Petrov2004}:
\begin{align}
a=\frac{mg^2}{4\pi}\frac1{\frac{mg^2\Lambda}{2\pi^2}-\nu_0},
\hspace{1cm} R^*=\frac{4\pi}{m^2g^2}.
\end{align}
The range parameter $R^*$ is necessary to fix the size of the smallest (i.e., ground-state) Efimov trimer consisting of two $\ket{1}$ atoms and one $\ket{2}$ atom. Note that previous experimental studies of identical bosons have found that the size of the ground-state Efimov trimer is  universally related to the van der Waals range \cite{Berninger2011}, an effect which was explained in Ref.~\cite{Wang2012}. Hence, it is natural to fix the Efimov physics using two-body parameters. From an investigation of the  vacuum three-body problem within the two-channel model, we find that the scattering length at which the ground-state Efimov trimer crosses the three-atom continuum threshold is $a_-\simeq-5000R^*$ \cite{Levinsen2015}. Since $R^*=60a_0$ in our experiment, we find that $|a_-|=3\times10^5a_0$, which exceeds the average interparticle spacing by two orders of magnitude. Thus, we expect Efimov physics to play a very small role in the experimental results. We emphasise that this conclusion is independent of the specific manner in which we include Efimov physics; indeed, calculations using realistic interatomic potentials find a similarly large separation of scales between the van der Waals range and $a_-$ \cite{Wang2012_2}.

Note that we do not apply the commonly used Fr{\" o}hlich approximation to the Hamiltonian for the impurity problem, as this would not allow us to consider near resonant interactions. Indeed, the Fr{\" o}hlich model already misses terms at the third order of perturbation theory~\cite{Christensen2015}. These are on the other hand correctly captured within the variational approach described in Ref.~\cite{Levinsen2015}, which forms the basis of our evaluation of the entire impurity spectral function.
 
\subsection{Truncated basis method for the Bose polaron}
To approximately model the Bose polaron across the full range of impurity-boson interactions, we apply a truncated basis method (TBM), first introduced in Ref.~\cite{Cetina2016}. This method was successfully used to model both the dynamics and the spectral response for an impurity strongly interacting with a Fermi gas~\cite{Cetina2016}, and here we extend the TBM to obtain the spectral function of the Bose polaron.

The TBM consists in truncating the Hilbert space of wavefunctions for the impurity in the BEC. In the present work, we restrict the Hilbert space to wavefunctions containing the impurity, the BEC, and up to 2 Bogoliubov excitations of the BEC. As we shall see, this allows us to capture the attractive and repulsive polaron peaks in the spectral function, as well as the continuum of states in between. Variational wavefunctions with up to one~\cite{Li2014} or two~\cite{Levinsen2015} Bogoliubov excitations have already been successfully used to determine the ground-state energy of the Bose polaron; here we extend the use of the variational wavefunction in Ref.~\cite{Levinsen2015} to evaluate the entire spectral function of the impurity. 

We start by considering an exact energy eigenstate of the system which satisfies the equation:
\begin{align} \label{eq:ham}
{\hat H} \ket{\psi} = E \ket{\psi}
\end{align}
where $E$ is the energy of the state $\ket{\psi}$. We then take truncated wavefunctions of the form:
$\ket{\psi} = \sum_j \alpha_j \ket{j}$, where $\{\ket{j} \}$
represents a subset of a complete orthonormal set of states. Inserting this into Eq.~\eqref{eq:ham} and taking the projection with respect to $\ket{j}$ then yields
\begin{align}\label{eq:trunc}
E \alpha_j = \sum_l  \bra{j} \hat{H} \ket{l} \alpha_l \equiv \sum_lH_{jl} \alpha_l.
\end{align}
Diagonalising the Hamiltonian within this subspace then corresponds to determining the eigenvalues and eigenvectors of the matrix $H_{jl}$. 

To determine the spectral function using the TBM, we exploit the relation between the Green's function in time and frequency space:
\begin{align}
A(\omega) & 
=2 {\rm Re} \int^\infty_0 dt \ e^{i\omega t}
            \bra{\psi_0} e^{-i\hat{H}t} \ket{\psi_0} 
= 2\pi \sum_{j} \left|\left<\psi_0 | \phi_j \right>\right|^2 \underbrace{\int^\infty_{-\infty} \frac{dt}{2\pi} e^{i\omega t} e^{-iE_jt}} _{\delta(\omega-E_j)} \ ,
\label{ATBM}
\end{align}
where $\ket{\psi_0}$ is the polaron state in the absence of interactions. Here, $\phi_j$ are the eigenstates of the truncated Hamiltonian, with energies $E_j$. In practice, the RF pulse is of finite duration in experiment, thus giving rise to a broadening of the measured spectrum. We model this non-zero Fourier width of the RF probe by convolving the spectral function with a Gaussian:
\begin{align} \label{eq:convolve}
I_0(\omega) & \equiv \int \frac{d\omega'}{2\pi}
                      A(\omega-\omega')\frac{1}{\sqrt{2\pi}\sigma_{\rm rf}}
                      e^{-\omega'^2/2\sigma_{\rm rf}^2}
\end{align}
where $\sigma_{\rm rf}$ corresponds to the Fourier width. Using \eqref{eq:convolve} in \eqref{ATBM} yields
\begin{align}
I_0(\omega) & 
= \sum_{j} \left|\left<\psi_0 | \phi_j \right>\right|^2 \frac{1}{\sqrt{2\pi}\sigma_{\rm rf}} e^{-(\omega - E_j)^2/2\sigma_{\rm rf}^2}.
\end{align}

For the specific case of an impurity in a BEC, we evaluate the spectrum using wavefunctions of the form
\begin{align}
  \ket{\psi} = \,  \Bigg( &\alpha_0 a^\dag_{\0,2} +  
                  \sum_\k \alpha_\k a^\dag_{- \k,2}
                  \beta^\dag_\k
+\frac{1}{2} \sum_{\k_1 \k_2} \alpha_{\k_1 \k_2} a^\dag_{- \k_1 - \k_2,2} 
                  \beta^\dag_{\k_1}   \beta^\dag_{\k_2} 
 +\gamma_0 d^\dag_\0 + 
                  \sum_\k \gamma_\k d^\dag_{- \k} \beta^\dag_\k 
                  \Bigg) \ket{\Phi},
\label{eq:psi}
\end{align}
with $\ket{\Phi}$ the wavefunction of the weakly interacting BEC. Such a wavefunction was first introduced in Ref.~\cite{Levinsen2015}, and provided the first many-body theory that included Efimov physics in a BEC. In this work, we include three-body correlations non-perturbatively in the impurity spectral function for the first time.

\subsection{Weak-coupling perturbation theory}
In the weak coupling limit, the polaron properties can be calculated perturbatively~\cite{Christensen2015}. Assuming $|a| \ll |a_-|$, the small parameter in this perturbative expansion is $a/\xi$, where we have defined the BEC coherence length $\xi\equiv1/\sqrt{8\pi n_0 a_B}$. To third order in $a$, the quasiparticle energy was calculated in Ref.~\cite{Christensen2015} to be 
\begin{align}
	\frac{E}{E_{n}}
		&=
			\frac{4}{3\pi} k_{n}\aI
				\Bigl[
			1 + \frac{8\sqrt{2}}{3\pi}
					\sqrt{\frac{4 k_{n} \aB}{3\pi}} k_{n}\aI
			  + \bigl( \frac{2}{3}-\frac{\sqrt{3}}{\pi} \bigr)
			  		\frac{4 k_{n} \aB}{3\pi} (k_{n}\aI)^{2}
			  	\ln\bigl(
			  		\sqrt{\frac{4 k_{n} \aB}{3\pi}} k_{n}a 
				  	\bigr)
			  	\Bigr].
	\label{eqn:E_pert}
\end{align}
When comparing perturbation theory for the polaron energy with the experimental data in Fig.\ 3 in the main manuscript, we  plot \eqref{eqn:E_pert} using a $k_n$ obtained from the trap averaged density. In principle, we should average \eqref{eqn:E_pert} over the cloud, but the difference between the results of the two averaging procedures is 
negligible in the perturbative regime. 

\begin{figure}[tb]
	\centering
		\centering
		\includegraphics{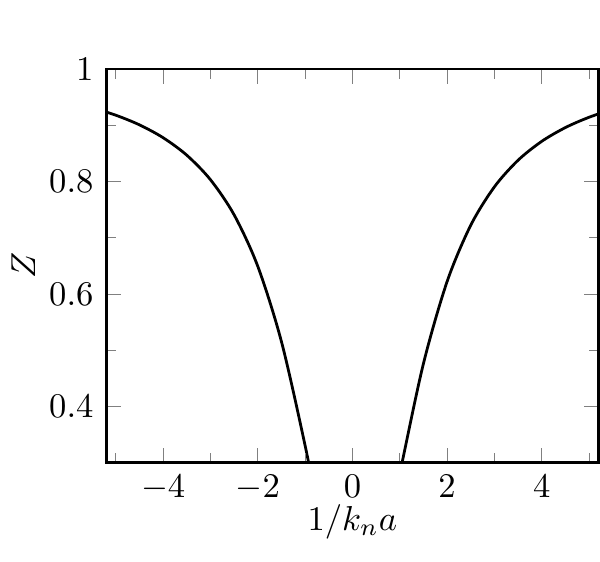}
		\caption{
			Quasiparticle residue of the attractive and repulsive polarons in a uniform BEC calculated from the perturbative expression Eq.\ (\ref{eqn:Z_pert}). 
			\vspace{1\baselineskip} }
		\label{fig:Quasiparticle_Residue}
\end{figure}

Likewise, the quasiparticle residue of the polaron to third order
in $a/\xi$~\cite{Christensen2015} is 
\begin{align}
	Z^{-1} &=
		1 + \frac{2\sqrt{2}}{3\pi}
				\frac{2}{\sqrt{3\pi k_{n} \aB}} (k_{n}\aI)^{2}
			+ 0.64 \times \frac{4}{3\pi} (k_{n}\aI)^{3}.
	\label{eqn:Z_pert}
\end{align}
Equations (\ref{eqn:E_pert})-(\ref{eqn:Z_pert}) together  determine the quasiparticle part of the impurity spectral function, $2\pi Z\delta(\omega-E)$, in the perturbative regime. The quasiparticle peak is dominant when  $1-Z\ll 1$, and it follows from Eq.\ (\ref{eqn:Z_pert}) that this condition corresponds to requiring $(k_n a)^2/\sqrt{k_n a_B}\ll 1$. Since our experimental value $k_n\aB\simeq 0.01$ is very  small, this condition is, in fact, stricter than the condition $\aI/\xi\ll1$. This means that the quasiparticle residue becomes significantly smaller than one, even when $\aI/\xi$ is still small.  Indeed, we see from  Fig.~\ref{fig:Quasiparticle_Residue}, that $Z \geq2/3$ only for $1/k_{n}\aI < -1.8$ or $1/k_{n}\aI > 2$.

In addition to the quasiparticle peak, the perturbative calculation can also provide some insight into  the broad many-body continuum part of the spectral function.
By expanding the self energy of the impurity up to second order in $a$, we obtain
\begin{align}
	A(\omega)
		&=
			2\pi Z\delta(\omega-E)
			+
			\theta(\omega) \frac{2}{E_{n}}
				\frac{
					\frac{2\sqrt{2}(k_{n}\aI)^{2}}{3\pi}
				\frac{[ \omega/E_{n} ]^{3}}
					{( 4k_{n}\aB/3\pi + \omega/E_{n} )^{5/2}}
				}
				{
				\frac{\omega^2}{E_{n}^2}
				+
				\Bigl(
				\frac{2\sqrt{2}(k_{n}\aI)^{2}}{3\pi}
				\frac{[ \omega/E_{n} ]^{3}}
					{( 4k_{n}\aB/3\pi + \omega/E_{n} )^{5/2}}
				\Bigr)^{2}
				},
	\label{eqn:spec_func_2nd}
\end{align}
where $\theta(x)$ is the Heaviside step function, and $E$ is the polaron energy in Eq.\ (\ref{eqn:E_pert}) up to second order. This result  illustrates the typical  shape of the impurity spectral function consisting of a  quasiparticle peak and a many-body continuum. It furthermore provides a simple physical interpretation of the continuum above the polaron energy for weak interactions: It consists of states formed by a Bogoliubov mode and the impurity moving with opposite momenta. The energy threshold of this continuum is zero within second order perturbation theory, because this is the minimum cost to create an impurity particle with momentum $\mathbf q$ and energy $q^2/2m$ plus a Bogoliubov mode with momentum $-\mathbf q$ and energy $E_\q$. However, on physical grounds, this continuum of states necessarily starts instead at the polaron energy, since Bogoliubov modes can be excited with arbitrarily small energy and momentum. For  large energy $\omega\gg 4\pi \aB n_0/m$, the Bogoliubov modes become ideal gas excitations with energy $\sim q^2/2m$, and the weight of the continuum spectrum of $A(\omega)$  decreases as $\omega^{-3/2}$. 
 
 \subsection{Comparison of spectral functions for the uniform system}
\begin{figure}[tb]
	\centering
	\begin{minipage}{.49\textwidth}
		\centering
	\includegraphics{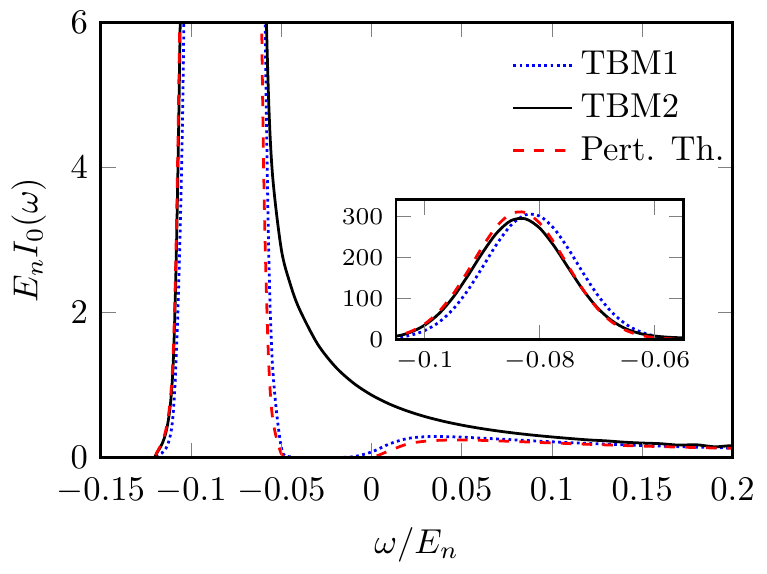}
	\end{minipage}
	\begin{minipage}{.49\textwidth}
		\centering
	\includegraphics{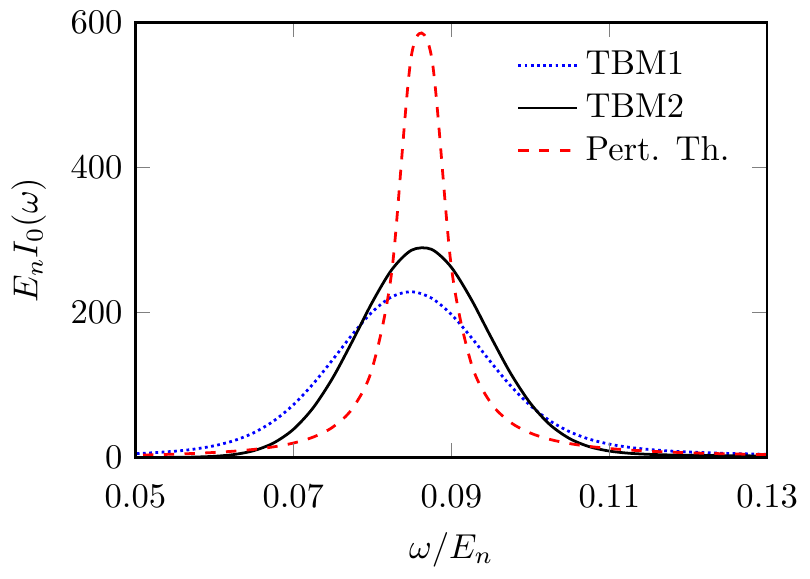}
	\end{minipage}
	\caption{ Spectral function for the impurity in a uniform BEC with $1/k_{n}\aI = -5$ (left) and $1/k_{n}\aI = 5$ (right) including a small Gaussian broadening $\sigma_{\textnormal{rf}}/E_{n}= 0.008$. The dashed line is the result of perturbation theory, the dotted line the TBM including only one Bogoliubov excitation, and the solid line the TBM including two Bogoliubov excitations. The inset shows the polaron peak for $1/k_{n}\aI = -5$.}
	\label{fig:spec_func_comp_weak_repulsive}
	\label{fig:spec_func_comp_weak_attractive}
\end{figure}

In Fig.\ \ref{fig:spec_func_comp_weak_attractive}, we plot the spectral function, convoluted with a small Fourier width according to \eqref{eq:convolve}, in the weak coupling regime $1/k_n\aI=\pm5$. We have used three different calculations to obtain $A(\omega)$: perturbation theory given by \eqref{eqn:spec_func_2nd}, the truncated basis method with only one Bogoliubov excitation included (TBM1), i.e., neglecting the third and fifth terms in Eq.\ (\ref{eq:psi}), and the full calculation including two Bogoliubov excitations (TBM2).  First, we see that all three calculations agree very well concerning the peak position, which corresponds to the energy of the polaron quasiparticle. This shows that the TBM recovers the perturbative result in the weak-coupling limit, as desired.

\begin{figure}
		\centering
	\includegraphics{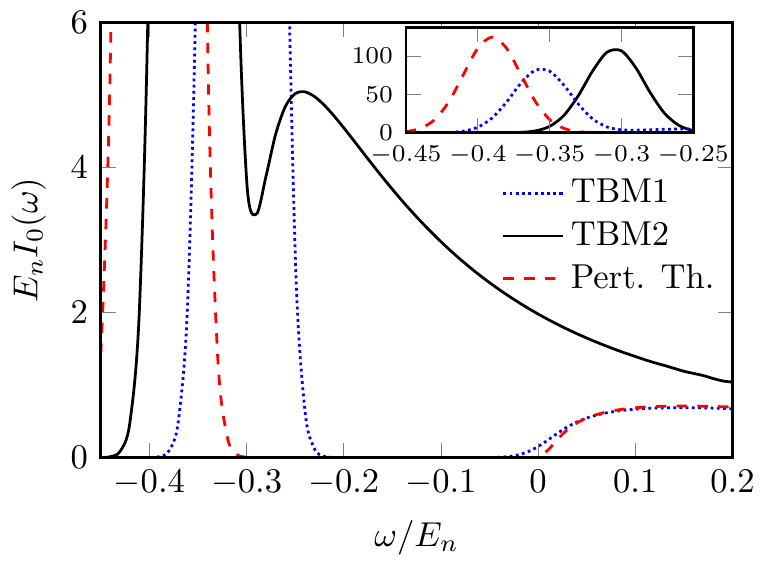}
	\caption{
	 Calculated spectral function for the impurity in a uniform BEC with $1/k_{n}\aI = -1$ and a small  Gaussian broadening $\sigma_{\textnormal{rf}}/E_{n}= 0.020$. Lines and inset are as in Fig.\ \ref{fig:spec_func_comp_weak_attractive}.}
	\label{spec_func_comp_strong_attractive}
\end{figure}

For the attractive case, we also see that both the perturbative calculation and TBM1 predict essentially the same many-body continuum, which  starts at zero energy as discussed above. TBM2, on the other hand, correctly predicts the continuum to start above the polaron peak (the transition from the polaron peak to the continuum is  smoothened due to the small Fourier broadening). This is because the wavefunctions with an extra Bogoliubov mode can describe dressed impurities at finite momentum.

In Fig.\ \ref{spec_func_comp_strong_attractive}, we plot the Fourier broadened spectral function for  $k_n\aI=-1$ obtained again from the three different calculations. For this fairly strong coupling, there is poor agreement between perturbation theory and the TBM, as expected. The three calculations give different predictions for the polaron energy as well as for the many-body continuum. In this regime, perturbation theory is not accurate and the TBM with two Bogoliubov modes is the most reliable, since it includes up to 3-body correlations non-perturbatively.

\section{Trap averaged spectra}

The preceding analysis was for an impurity in a BEC of uniform density $n_0$. However, in the experiment, the atomic BEC is confined in a harmonic trap, and instead has \textit{average} density $n_0$, with corresponding $k_n$. On the scale of the trap, the RF probe is essentially uniform, transferring atoms from the $\ket{1}$ state into the $\ket{2}$ state in all regions of the inhomogeneous BEC. This in turn gives rise to a broadening of the observed spectral response of the impurity atom, since it is surrounded by a BEC of varying density $n(\mathbf r)$. We take this into account using the local density approximation to average the response over the cloud:
 \begin{align}
 I(\omega) = \frac{1}{N}\int d^3\mathbf{r} \ n(\mathbf{r}) I_0(\omega,n(\mathbf{r})).
 \label{RFtrapaveraged}
\end{align}
Here $I_0(\omega,n(\mathbf{r}))$ is the local Fourier broadened response obtained from \eqref{eq:convolve}  using an impurity spectral function $A(\omega)$ corresponding to the  density $n(\mathbf{r}) = \frac{m}{4\pi\hbar^2 a_{bb}} \mu(\mathbf{r})$ with $\mu(\mathbf{r})=\mu-V_\text{trap}(\mathbf{r})$.

\begin{figure}
		\centering
	\includegraphics[width=16.5cm]{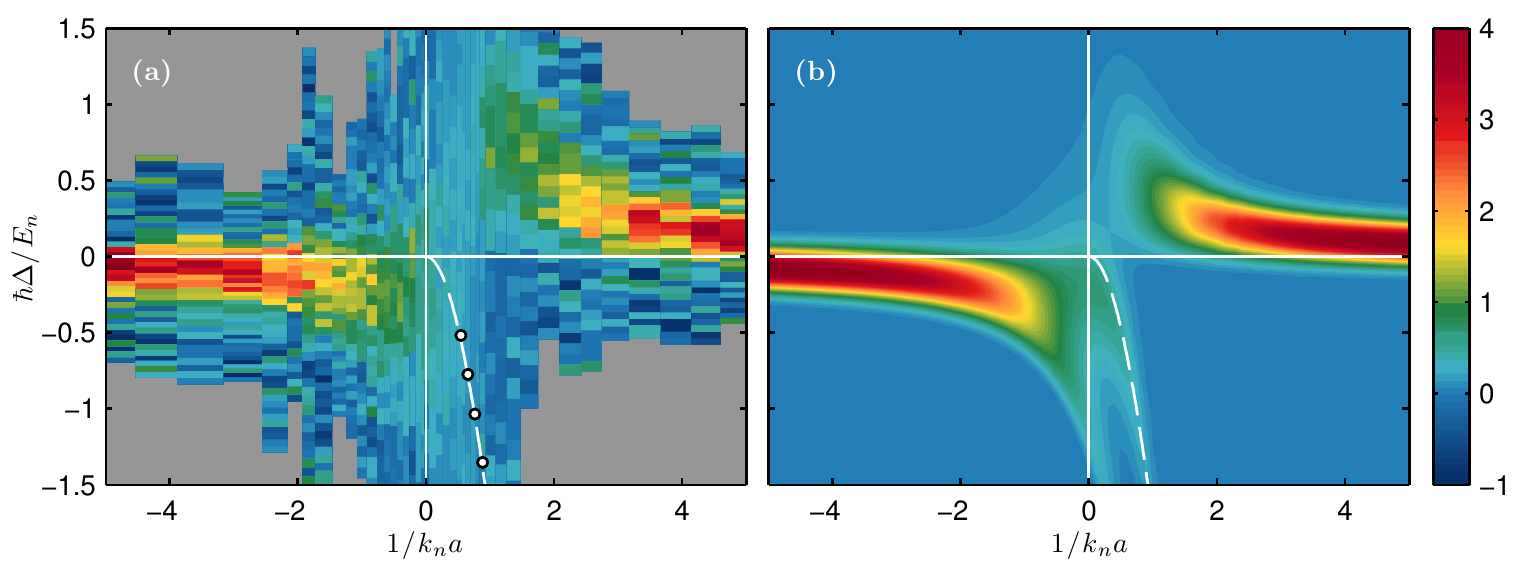}
	\caption{The experimental spectrum (a) and the trap-averaged spectral function calculated within TBM (b), both normalised so that the frequency integrated weight is the same for each interaction strength. The experimental Fourier width is estimated to be $\sigma_{\rm rf} \simeq 0.08 E_n$.}
	\label{fig:spectbm}
\end{figure}
 
For weak interactions,  the spectral function is dominated by the quasiparticle peak, such that Eq.~\eqref{RFtrapaveraged} becomes
\begin{align}
	I(\omega)
		& \simeq
			\frac{1}{N} \int\! \voldif{r}\, n(\bm{r})
				\frac{1}{\sqrt{2\pi\sigma_{\rm rf}^{2}}}
					e^{\frac{( \omega-E(\bm{r}) )^{2}}{2\sigma_{\rm rf}^{2}}}.
					\label{RFbroadened}
\end{align}
where $E(\mathbf{r})$ is the local quasiparticle energy at position $\mathbf{r}$ in the trap. This is the expression used to calculate the perturbative spectral width in Fig.~4 of the main manuscript. In the limit $|k_na| \ll 1$, the width of the spectral signal is dominated by the Fourier width $\sigma_{\rm rf}$ of the RF probe, since the trap averaging only occurs over a small range of local interaction parameters. However, for stronger coupling, the signal is averaged over an increasingly larger range of local interaction parameters and thus the trap inhomogeneity can significantly broaden the quasiparticle peak. In this regime, the many-body continuum is also modified by the trap.

The full trap-averaged spectral function obtained within TBM is shown in Fig.~\ref{fig:spectbm} together with the experimental result. The exact same data is presented in Fig.~2 of the main text, but here we normalise so that the frequency integrated weight is the same for each interaction strength. This illustrates more clearly how the spectral weight of the many-body continuum suppresses that of the quasiparticle peak in the strongly interacting unitary region.

\bibliography{BosePolRef}

\begin{thebibliography}{54}
\bibitem [{foo()}]{footnote1}%
  \BibitemOpen
  \href@noop {} {}\bibinfo {note} {At weak and moderate interactions, the
  signal obtained from the thermal atoms confirms the position of the bare
  transition $\omega_0$ which is obtained from independent
  measurements.}\BibitemShut {Stop}%
\end{thebibliography}

\end{document}